\documentclass[prd,
notitlepage,
11pt,aps,
nofootinbib,%twocolumn,
floatfix%,superscriptaddress
]{revtex4}
\usepackage{graphicx}
\usepackage{epsf,amsmath,amsfonts,amssymb,amsbsy}
\usepackage[mathscr]{eucal}

\begin{document}

\title{Genesis of spiral galaxies}

\author{Valery V. Kiselev}\email{Valery.Kiselev@ihep.ru}
\affiliation{Russian State Research Center Institute for High Energy Physics
(National Research Center Kurchatov Institute), Russia, 142281, Moscow
Region, Protvino, Nauki 1} \affiliation{Moscow Institute of Physics and
Technology (State University), Russia, 141701, Moscow Region, Dolgoprudny,
Institutsky 9}

\begin{abstract}
Enigmatic spiral structure of many galaxies and its huge orbital momentum
originated due to the capture of lightweight bare black hole by gravity of
heavy primordial gas cloud at large impact parameter. The rotating of black
hole caused the formation of accretion disc from the cloud and the transfer
of orbital momentum to the disc, while during the fall to the center of mass,
the spiral trace of black hole in the disc did create the spiral front line
of sound waves in the gas, that further evolved into the stellar spiral arms.
This mechanism opens the way to study features of spiral galaxy formation,
say, an influence and a significance of dark matter in this process.
\end{abstract}

%\keywords{galaxies: spiral, structure}
\maketitle

\section{Introduction}

Spiral galaxies refer to the main class of star clusters coupled by gravity.
The structure of such galaxies is very peculiar and amazingly fine-tuned
\cite{1}:
\begin{itemize}
  \item stars rotate around the center to the same direction, that
      corresponds to a huge initial orbital momentum of unknown origin,
      in contrast to elliptic galaxies, wherein the stellar motion is
      chaotic in direction and mainly radial with a little rotation,
  \item outside the central bulge the galaxy is flat, where the
      distribution of stars has dense regions of nontrivial geometry in
      the form of spiral arms,
  \item the spiral arms contain bright young big stars, while the bulge
      and voids in the galaxy disk are populated by low-mass old stars,
  \item often, the central region contains the bar connecting two or more
      spirals,
  \item moreover, if there are observational data, astronomers certainly
      find a supermassive black hole posed in the galaxy center
      \cite{4,5}, and it is not clear, whether such black hole does
      locate in each spiral galaxy or not.
\end{itemize}
Despite of long interest to study these problems \cite{6,7,8}, all of above
issues do not yet find a complete overall explanation. We establish that all
of those features are caused by the single mechanism: the capture of bare
black hole by the cloud of primordial gas.

\section{The very begin: capture and disc}

During the Universe evolution, spatially inhomogeneous primordial gas gets
cold in the form of heavy clouds contracted by the gravity. We suppose that
at this stage there is comparable number of bare lightweight black holes,
dense close to the same seeds of inhomogeneity. It would be natural to get
the maximal probability that the cloud and black hole have a relative motion
with a large impact parameter, greater than or comparable to the size of
cloud. If the relative velocity of black hole is less than a critical value,
then it would be captured by the gravity of gas cloud. For the sake of
simplicity of qualitative picture, we suppose that the black hole begins to
rotate around the cloud on the circular orbit far away of dense region of the
cloud. If the orbit is elliptical, the black hole can enter to the more dense
regions in the cloud periphery that result in the growth of black hole mass
at the loss of velocity and its further capture close to the center of mass
of system without the transfer of orbital momentum to the gas cloud. However,
the elliptical orbit can be beyond the dense regions of the gas, so that the
black hole will rotate similar to the case of circular orbit.

The rotation takes place around the center of mass posed in vicinity of
center of massive gas cloud, of course. It could be long enough for that the
gravity of black hole attracts the gas from the cloud in the plane of orbit
that gives the formation of gas disc between the black hole and the cloud.

This disc rotates in the same direction as the black hole. This transfer of
orbital momentum is possible, if the black hole is slightly displaced close
to the center of mass. A little fraction of disc in vicinity of black hole is
absorbed by the black hole that means the accretion, the growth of black hole
mass. The driftage of black hole provides the formation of flat structure
around the primordial gas cloud and the orbital momentum transfer to the
accretion disc (see Fig. 1).

\begin{figure*}[th]
\centerline{
  \includegraphics[width=5cm]{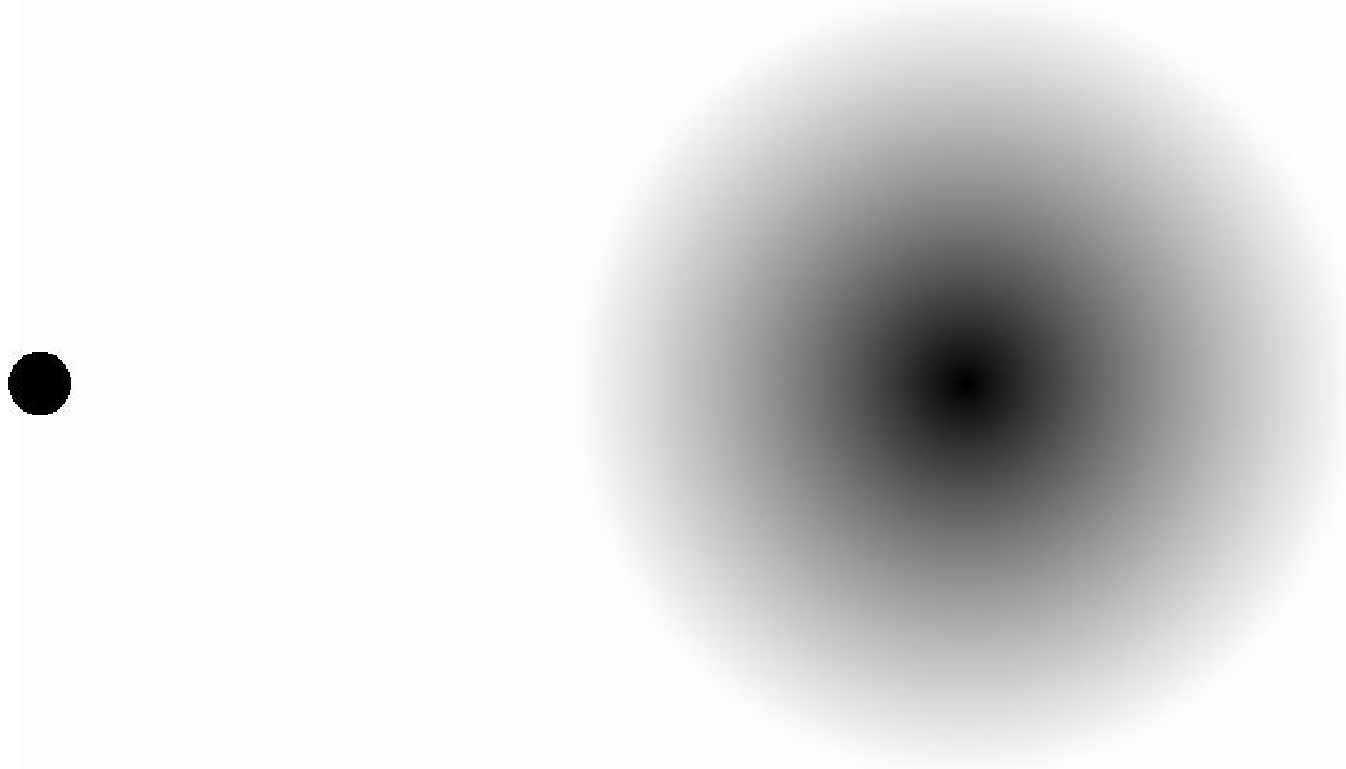}\hspace{3cm}
  \includegraphics[width=5cm]{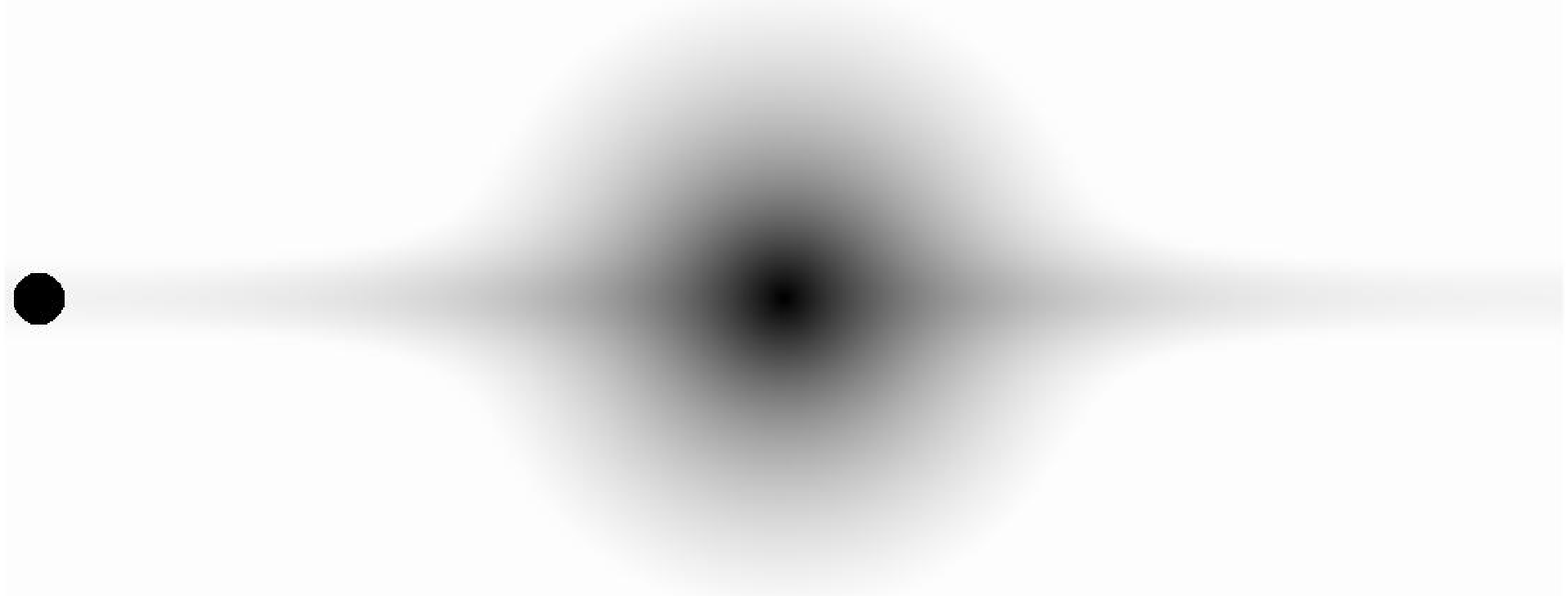}
  }%\\
  \caption{The bare black hole captured by the cloud of primordial gas
  before and after the creation of accretion disc (left and right, respectively).}\label{f1}
\end{figure*}

\section{Spiral trace and waves}

When the growth of accretion disc passes a critical value of density, the
driftage of black hole ends and the accretion gets the avalanche character,
so that the black hole rapidly increases in mass and falls down to the center
of mass. This fast fall with rotation means the spiral form of trajectory,
indeed, as pictured in Fig. 2 (left).

\begin{figure*}[th]
\centerline{
  \includegraphics[width=5cm]{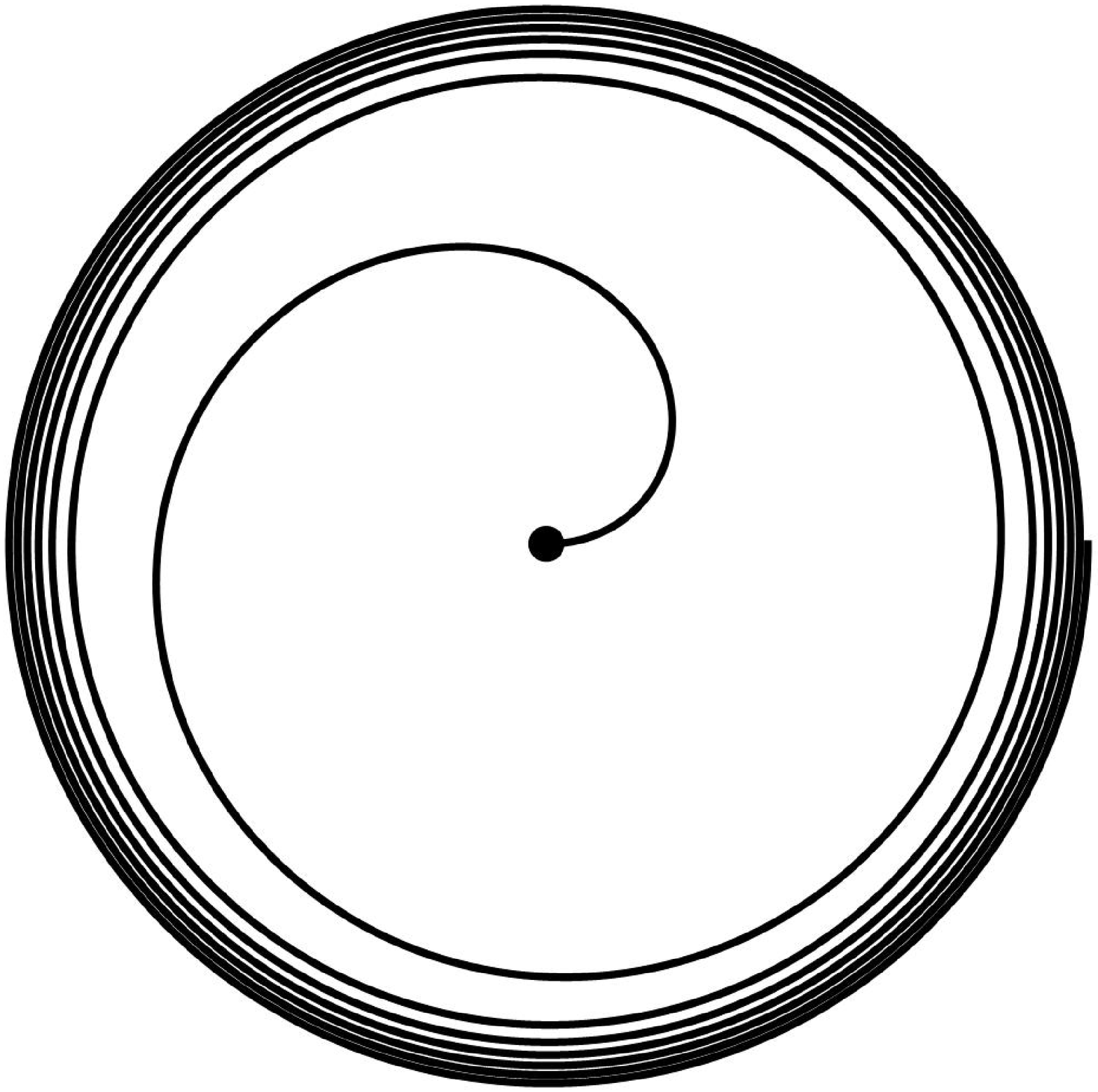}\hspace{3cm}
  \includegraphics[width=5cm]{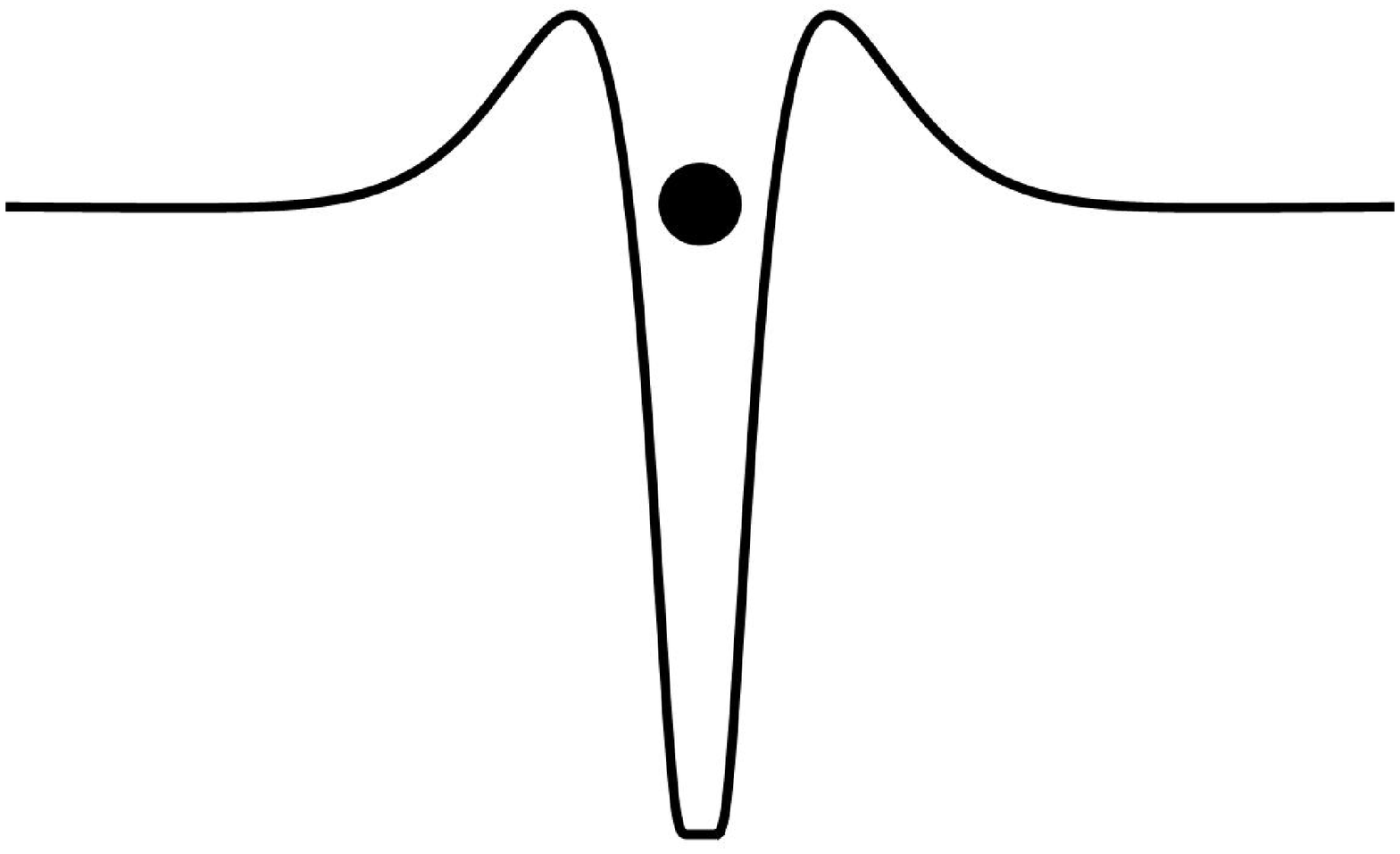}
}%\\
  \caption{The driftage of black hole with further fall down
  to the center (left, the view to the plane of rotation)
  and the profile of gas density in the direction perpendicular
  to the trajectory of black hole in the plane just behind
  the black hole (right).}\label{f2}
\end{figure*}

The trace of the spiral trajectory in the gas gets the specific profile of
gas density: the black hole orbit itself remains the under-dense curve
because of the absorption of gas, while the very vicinity of this curve is
over-dense because of the attraction by the black hole during the accretion.
This profile of gas density in the transverse direction to the black hole
trajectory is shown in Fig. 2 (right). The inhomogeneity of gas starts to
propagate as the sound. It is important that the front line of sonic wave is
spiral. Since the gas is rather cold the speed of sound should be small, so
that several spiral waves can form. Mostly probable, the speed of disk
rotation exceeds the speed of sound, so that two dominant spiral waves move
in the same direction as the disk rotation, with the difference of velocities
equal to the double sound speed. The magnitude of density gradient can be
significant in order to produce shock waves, which can possess some
properties of soliton, i.e. the wave with no damping. Sub-leading sound waves
or harmonics can follow the dominant primary spiral front lines. Further, the
gas inhomogeneity will be contracted because of the self-gravitation during
the star formation. The dense regions are transfered to the stellar spiral
arms in the flat spiral galaxy possessing the huge orbital momentum.

\section{Stars, bars and black hole mass}

Note that in general, the more rapid rotation of disk differs from the slow
rotation of bulge. Moreover, the sound waves correspond to the arranged
collective motion of dense worm gas hence it should delay the process of
inhomogeneity growth. Therefore, the star formation should be rather
different in the two regions: the dense hotter spiral arms versus the colder
bulge and under-dense disc voids. Then, the arms should contain more young
and big stars in contrast to the bulge and voids containing more old and less
massive stars as observed, indeed. The dense gas in spiral arms still
activates the star formation, while bright massive stars have smaller time of
life and they can leave the dense regions and evolve to more dark voids.

In addition, the shock sound waves created by the black hole in the discIn
addition, the shock sound waves created by the black hole in the disc would
differ from those waves in the more dense and slow bulge, even the speed of
sound would be different. The accretion of matter to the black hole is more
intensive in the bulge, while the loss of rotation velocity is rapid, so that
the black hole should almost stop the rotation in the bulge and move
straightforwardly to the center of mass. This fact means that the spiral form
can be transformed into the straight line in the region of dense bulge. Thus,
in many cases the two spirals are accompanied by the bars in the center of
galaxy. Initially, these two arms are close to each other and begin to
diverge. From the symmetry and stabilization arguments, we deduce that the
most probable final location of two arms would be opposite to each other, so
that the two bars in the center of bulge should form the single common bar
connecting one spiral to the other. Moreover, the stars formed in the bulge
bar can move on elliptical orbits with large eccentricities. Since the
rotations of galaxy disc and its bulge are different, generically, there is
no exact synchronization of the rotations for the spiral and its bar, so that
the simple point-like connection between the spiral and its bar can be
broken. Empirically, two thirds of spiral galaxies population contains the
bar.

Finally, the mechanism requires that all of spiral galaxies have to contain
the supermassive black holes in the centers. In addition, the bare mass of
black hole is deeply overlapped by the mass of accretion \cite{10,11}, which
directly depends on the mass of primordial gas cloud. Therefore, the mass of
super-heavy black hole has to correlate with the visible mass of galaxy, that
seems to be confirmed empirically \cite{K,M1,M2,G}.

\section{Discussion}

This mechanism of genesis for spirals is clearly justified and fits to
observed features under the known laws of physics. It can be compared with
other realistic models treating the regularities in spiral galaxies.

In the first approach \cite{L1,L2,LS}, the spirals are assigned to the
density waves appearing due to instabilities in the rotation of central
region. We see that this approach accepts the rotation as the given condition
with no explanation of its origin, while the main idea about the density wave
nature of spirals is incorporated in the mechanism described in the present
paper.

The second model \cite{Shock1,Shock2,Shock3} considers the star formation in
the interstellar medium because of shock waves created by stellar winds and
explosions of supernovae. Fronts of shock waves can take the spiral form,
too. So, the shock wave idea is also incorporated in our mechanism, but in
contrast, the spiral fronts are certainly created by the passage of black
hole through the interstellar medium. Again, the second approach assumes the
rotation of disc galaxy as the given fact.

So, we see that advantages of previous developments in the field are included
in the present model and it is enforced by important ingredients implementing
the origin for the rotation and its huge orbital momentum as well as for the
flat disc and spiral fronts of seeds for the shock waves.

In addition, we mention studies of hydro-dynamical analog for the spiral
galaxies \cite{Fr} and fresh results in computer simulations of spiral
structure formation in rotating disc of gravitating medium \cite{comp}.

Thus, the suggested mechanism is the first that draws the complete clear
pattern for the genesis of spiral galaxies and now it can be used for the
investigation of features in the process of formation of spiral galaxies. For
instance, one could study a role played by the dark matter \cite{24,25},
which has to dominantly accompany any dense inhomogeneity of ordinary matter.
Such investigations are complicated, since we need to account for the balance
of kinetic and potential energy for each microscopic element of primordial
cloud of matter in order to make reliable calculations of evolution, but they
could clarify our understanding of visible spectrum of properties in spiral
galaxies, on the base of rigorous dynamical foundations. Another question
concerns for a justification of the main assumption in this paper: What is
the mechanism of bare black hole creation, and how do we estimate the
population of bare black holes? Therefore, the conjecture needs further
cumbersome computer simulations as well as a progress in the early history of
black holes in the Universe.

\acknowledgments

The author thanks profs. A.K.Likhoded and G.P.Pron'ko for discussions.

\end{document}